\documentclass[aps,pra,reprint, amsmath, amssymb,superscriptaddress,nofootinbib]{revtex4-1}

\usepackage{bm}
\usepackage[retainorgcmds]{IEEEtrantools}
\usepackage{graphicx}
\usepackage{mathrsfs}
\usepackage{amsmath}
\usepackage{amssymb}
\usepackage{color}
\usepackage{amsfonts}
\usepackage{times,txfonts}
\usepackage{nicefrac}
\usepackage[colorlinks=true,linkcolor=blue,urlcolor=blue,citecolor=blue,pdfusetitle]{hyperref}
\usepackage{amsmath}
\DeclareMathOperator{\sign}{sign}

\newcommand{\tr}{\text{tr}}

\begin{document}

\title{Detailed fluctuation theorem bounds apparent violations of the second law}
\date{\today}
\author{Domingos S. P. Salazar}
\affiliation{Unidade de Educa\c c\~ao a Dist\^ancia e Tecnologia,
Universidade Federal Rural de Pernambuco,
52171-900 Recife, Pernambuco, Brazil}

\begin{abstract}
The second law of thermodynamics is a statement about the statistics of the entropy production, $\langle \Sigma \rangle \geq 0$. For small systems, it is known that the entropy production is a random variable and negative values ($\Sigma < 0$) might be observed in some experiments. This situation is sometimes called apparent violation of the second law. In this sense, how often is the second law violated? For a given average $\langle \Sigma \rangle $, we show that the strong detailed fluctuation theorem implies a lower tight bound for the apparent violations of the second law. As applications, we verify that the bound is satisfied for the entropy produced in the heat exchange problem between two reservoirs mediated by a bosonic mode in the weak coupling approximation, a levitated nanoparticle and a classical particle in a box.
\end{abstract}
\maketitle{}


{\bf \emph{Introduction -}} 
In recent years, interest in nonequilibrium phenomena of small systems escalated \cite{RevModPhys.93.035008,Seifert2012,Campisi2011,Bustamante2005,Esposito2009}. The areas of stochastic and quantum thermodynamics took a new turn with the advent of the Fluctuation Theorem (FT) and its variants \cite{Jarzynskia2008,Jarzynski1997,Jarzynski2000,Crooks1998,Gallavotti1995,Evans1993,Hanggi2015,Saito2008,PhysRevX.11.031064}. In several situations, when probing small systems, the object of interest is the entropy production, $\Sigma$, now modelled as a random variable with relevant fluctuations in nonequilibrium setups. In this case, the FT gives additional information regarding the fluctuation  of the entropy production.

Particularly, the strong Detailed Fluctuation Theorem (DFT) is a relation about the asymmetry of the probability density function of the entropy production,  \begin{equation}
\label{DFT}
   \frac{p(\Sigma)}{p(-\Sigma)}=e^{\Sigma},
\end{equation}
indicating that positive values of entropy production are more likely to be observed than the negative counterparts. It arises, for instance, in time symmetric protocols in the exchange fluctuation framework \cite{Hasegawa2019,Timpanaro2019B,Evans2002,Merhav2010,Garcia2010,Cleuren2006,Seifert2005,Jarzynski2004a,Andrieux2009,Campisi2015}. The most known consequence of (\ref{DFT}) is the integral fluctuation theorem (IFT), $\langle e^{-\Sigma}\rangle=1$, which results in the second law of thermodynamics, $\langle \Sigma \rangle \geq 0$, from Jensen's inequality.

Actually, the second law is not the only statement about the statistics of $\Sigma$. As the DFT (\ref{DFT}) is stronger than the second law (DFT$\rightarrow\langle \Sigma \rangle \geq 0$), it also imposes other general consequences for the statistics of $\Sigma$ as well \cite{Merhav2010,Timpanaro2019B,Hasegawa2019,Neri2017,Pigolotti2017}.

Within this framework, a new class of bounds relating the signal to noise ratio of currents to the average entropy production was discovered and named Thermodynamic Uncertainty Relations (TURs) \cite{Barato2015,Gingrich2016,MacIeszczak2018,Polettini2017,Pietzonka2017,Hasegawa2019,Timpanaro2019B}. When written solely in terms of the statistics of $\Sigma$, they read $var(\Sigma)/\langle \Sigma \rangle^2 \geq t(\langle \Sigma \rangle)$, for some known function $t(x)$. Note that this is a second order statistics about the entropy production. Other results derived from the FT followed the same idea of the TUR. For instance, a tight bound for the mean entropy production $\langle \Sigma \rangle$ using the asymmetry of the marginal statistics of currents \cite{Campisi2021}. Another example is the information bound for $p(\Sigma)$ for a given $\langle \Sigma \rangle$ given in terms of a maximal distribution \cite{Salazar2021}.

In this context, the so called apparent violation of the second law is a rather simple statement about the statistics of $\Sigma$. Namely, it is given by the magnitude of $P(\Sigma < 0)$ \cite{Jarzynskia2008a,Merhav2010}.  Of course, it is called apparent because the second law is a statement about the average $\langle \Sigma \rangle\geq0$, so negative values in $\Sigma$ do not constitute an actual violation. There have been advances in placing a bound for the apparent violation using a variety of constraints. For instance, for the constrained domain, $\Sigma\leq\Sigma_{max}$, a bound for the apparent violation has been established using the IFT \cite{Cavina2016,Kantz2019} and implemented experimentally
\cite{Maillet2019}, originally written in terms of the irreversible work. For situations where the DFT is suitable, it is easy to check that (\ref{DFT}) immediately bounds the apparent violation,  $0\leq P(\Sigma<0) \leq 1/2$. Actually, the authors in \cite{Merhav2010} used the DFT and obtained a lower bound for the apparent violation in terms of the conditional average, $\langle \Sigma \rangle_+=\int_0^\infty \Sigma P(\Sigma)d\Sigma$.

However, in the same spirit of the TUR \cite{Timpanaro2019B}, using only the average as constraint $\langle \Sigma \rangle$,  the existing lower bounds \cite{Jarzynskia2008a,Kantz2019,Merhav2010} for the apparent violation from the DFT (\ref{DFT}) are currently loose to our knowledge.

In other words, for a given $\langle \Sigma \rangle$, is there a way to use the DFT (\ref{DFT}) and find a lower bound for such apparent violation? This is the question addressed in this paper. We show that, for a given $\langle \Sigma \rangle$, a system satisfying the DFT (\ref{DFT}) has the following tight and saturable lower bound for the apparent violation of the second law,
\begin{equation}
\label{introB}
   P(\Sigma<0)+\frac{P_0}{2}\geq\frac{1}{2}\Big(1-\frac{\langle \Sigma \rangle}{g(\langle \Sigma \rangle)}\Big),
\end{equation}
for $g(x)$ the inverse function of $h(x):=x\tanh(x/2)$ and $P(\Sigma<0)=\int_{-\infty}^{0}p(\Sigma)d\Sigma$ is the cumulative distribution. $P_0$ is the point mass function at $\Sigma=0$, $P_0=\lim_{\varepsilon \rightarrow 0}\int_{-\varepsilon}^\varepsilon p(\Sigma)d\Sigma$, usually zero for continuous $p(\Sigma)$, but it might be nonzero in some discrete cases (for instance, see the bosonic mode and qubit swap engine in the applications). Note that the rhs in (\ref{introB}) is given solely in terms of the mean $\langle \Sigma \rangle$. We show that the distribution that saturates (\ref{introB}) is the same minimal distribution that saturates the TUR \cite{Timpanaro2019B}. For completeness, we also prove that the intuitive upper bound, $P(\Sigma<0)+P_0/2 \leq 1/2$, is actually the best possible bound in this setup ($\langle \Sigma \rangle$ as the single constraint) by explicitly constructing a family of distributions $p'(\Sigma)$ that saturate this bound in the appropriate limit.

As applications, we show that the bounds are satisfied for the heat exchanged between two reservoirs mediated by three different systems: a bosonic mode in weak coupling, a classic levitated nanoparticle and a classic particle in a box. We also include the qubit swap engine and gaussian cases for comparison.

{\bf \emph{Formalism -}}
 We define the total variation (TV) distance between two probability density functions, $p(\Sigma)$ and $q(\Sigma)$, as $\Delta(p,q):=Sup_{A\subset\mathbb{R}} |P(A)-Q(A)|$, where $P(A):=\int_A p(\Sigma)d\Sigma$. Now we apply $\Delta$ to the pair $p(\Sigma)$ and $\hat{p}=p(-\Sigma)$. It can be written as
\begin{equation}
\label{TV1}
    \Delta(p,\hat{p})=\frac{1}{2}\int_{-\infty}^\infty |p(\Sigma)-p(-\Sigma)|d\Sigma. 
\end{equation}
Using the DFT (\ref{DFT}), we have $p(\Sigma)>p(-\Sigma)$ iff $\Sigma>0$, which simplifies (\ref{TV1}) to the expression
\begin{equation}
\label{TV2}
     \Delta(p,\hat{p})=\int_{-\infty}^\infty \sign(\Sigma)p(\Sigma)d\Sigma=\langle \sign(\Sigma)\rangle, 
\end{equation}
where $\sign(\Sigma)=\Sigma/|\Sigma|$, for $\Sigma \neq 0$ and $\sign(0)=0$. Unless stated otherwise, the averages $\langle . \rangle$ are taken over $p(\Sigma)$, $\langle . \rangle = \langle . \rangle_p$. Now we are going to obtain bounds for the total variation (\ref{TV3}) and connect it with the violation of the second law, here quantified as $P(\Sigma < 0)+P_0/2$.

\textit{Lower bound--}
For the lower bound, we consider the useful property of the average of odd functions under the DFT \cite{Hasegawa2019}. Let $u(\Sigma)=-u(-\Sigma)$ be an odd function, then
\begin{equation}
\label{oddproperty}
    \langle u(\Sigma) \rangle = \langle u(\Sigma) \tanh(\Sigma/2)\rangle,
\end{equation}
following directly from (\ref{DFT}). Using (\ref{oddproperty}) for the odd function $u(\Sigma):=\sign(\Sigma)$, we get from (\ref{TV2}):
\begin{equation}
\label{TV3}
    \Delta(p,\hat{p})=\langle \sign(\Sigma)\rangle = \langle \tanh(|\Sigma|/2)\rangle=\langle f(\Sigma)\rangle,
\end{equation}
where we used $\sign(\Sigma)\tanh(\Sigma/2)=\tanh(|\Sigma|/2):=f(\Sigma)$. Also from (\ref{oddproperty}), we have
\begin{equation}
\label{haverage}
    \langle \Sigma \rangle=\langle \Sigma \tanh(\Sigma/2)\rangle = \langle h(\Sigma)\rangle,
\end{equation}
for $h(\Sigma):=\Sigma \tanh(\Sigma/2)$.
Then, we define the inverse function, $g:=h^{-1}$, $g(h(\Sigma))=\Sigma$. We now use the standard procedure \cite{Y.Zhang2019,Campisi2021} based on Jensen's inequality from (\ref{TV3}):
\begin{equation}
\label{Jensens}
    \langle f(\Sigma)\rangle=\langle f(g(h)) \rangle \leq  f(g(\langle h(\Sigma) \rangle))=f(g(\langle \Sigma \rangle)),
\end{equation}
since $w(h):=f(g(h))$ results in $w'(h)>0$ and $w''(h)<0$, for $h>0$.
Comparing (\ref{Jensens}) and (\ref{TV3}), we obtain
\begin{equation}
    \label{TV4}
    \Delta(p,\hat{p})\leq \tanh(g(\langle \Sigma \rangle)/2)=\langle \Sigma \rangle/g(\langle \Sigma \rangle), 
\end{equation}
where we used $|g(x)|=g(x)$, for $x\geq0$ and $g(x)\tanh(g(x)/2)=h(g(x))=x$. 

Finally, note that from (\ref{TV2}) we get $\Delta(p,\hat{p})=P(\Sigma>0)-P(\Sigma<0)$. And from the normalization, $P(\Sigma >0)=1-P_0-P(\Sigma<0)$, it results in the lower bound for the apparent violation of the second law from (\ref{TV4}),
\begin{equation}
\label{lowerbound}
   P(\Sigma<0)+\frac{P_0}{2}=\frac{1}{2}\big(1-\Delta(p,\hat{p})\big)\geq\frac{1}{2}\big(1-\frac{\langle \Sigma \rangle}{g(\langle \Sigma \rangle)}\big).
\end{equation}
Note by inspection that the bound (\ref{lowerbound}) is achieved by the minimal distribution $p_a(\Sigma):=[e^{a/2}\delta(\Sigma-a)+e^{-a/2}\delta(\Sigma+a)]/(2\cosh(a/2))$, for $a=g(\langle \Sigma \rangle)$, the same distribution that appears in the context of the TUR \cite{Timpanaro2019B,Van_Vu_2020}. Also note that, for equilibrium, $\langle \Sigma \rangle = 0$, the rhs (\ref{lowerbound}) approaches $1/2$ and the DFT gives $P(\Sigma < 0) = P(\Sigma>0) \rightarrow P(\Sigma<0)=0$ and $P_0=1$ from the DFT. 

A final remark is that Pinsker's inequality, $\Delta(p,\hat{p})\leq \sqrt{D_{KL}(p|\hat{p})/2}=\sqrt{\langle \Sigma \rangle /2}$, gives a looser bound than (\ref{TV3}), where $D_{KL}(p|q)=\sum_i p_i \log (p_i/q_i)$ is the Kullback-Leibler divergence and $D_{KL}(p|\hat{p})=\langle \Sigma \rangle/2$ follows directly from (\ref{DFT}). One can easily show that $\langle \Sigma \rangle / g(\langle \Sigma \rangle) \leq \sqrt{\langle \Sigma \rangle/2}$ (where $g(x) \geq \sqrt{2x}$ for $x>0$ follows from $\kappa(x):=g(x)-\sqrt{2x}$ and $\kappa(x)'>0$, $\kappa(0)=0$). Actually, we have $g(x)\approx \sqrt{2x}$ for $x \approx 0$, which makes both bounds equivalent, $\langle \Sigma \rangle / g(\langle \Sigma \rangle) \approx \sqrt{\langle \Sigma \rangle/2}$, in the limit $\langle \Sigma \rangle\approx 0$.

\textit{Upper bound--}
It is immediate from the DFT that
\begin{equation}
\label{upperbound}
    P(\Sigma<0) + \frac{P_0}{2}\leq 1/2.
\end{equation}
For a fixed $\langle \Sigma \rangle$ and assuming (\ref{DFT}), can the bound (\ref{upperbound}) be improved? For this setup, the answer is no. One needs additional constraints to improve (\ref{upperbound}). 

We show that, for any given $\epsilon$, the inequality $P(\Sigma<0)+P_0/2\leq 1/2(1-\epsilon)$ is not satisfied for some $p'(\Sigma)$ with the same $\langle \Sigma \rangle$, which makes (\ref{upperbound}) the best upper bound in this setup. The proof goes as follows. Suppose we have a $\epsilon=\epsilon(\langle\Sigma\rangle)<\Delta^*$, where we used the upper bound (\ref{TV4}), $\Delta^*:=\tanh(g(\langle \Sigma \rangle)/2)$, such that 
\begin{equation}
\label{ineq}
    P(\Sigma<0)+\frac{P_0}{2}=\frac{1}{2}(1-\Delta)\leq\frac{1}{2}(1-\epsilon)\rightarrow \Delta \geq \epsilon,
\end{equation}
holds for all $p(\Sigma)$, where $\Delta=\Delta(p,\hat{p})$. It means the total variation (\ref{TV1}) would have a lower bound $\epsilon>0$. 

In order to show that (\ref{ineq}) is not true, our goal is to create a distribution $p'(\Sigma)$ such that DFT (\ref{DFT}) holds, it has a fixed average $\langle \Sigma \rangle$ and a total variation $\Delta'<\epsilon$. For that purpose, consider the ansatz:
\begin{equation}
\label{maxdist}
    p'(\Sigma)=\eta p_{a'}(\Sigma) + (1-\eta)\delta(\Sigma),
\end{equation}
for $p_{a'}$ minimal distribution with parameter $a'$, and $0\leq \eta \leq 1$, and $\delta(\Sigma)$ is a Dirac's delta function. Notice that (\ref{maxdist}) is normalized and satisfies the DFT (\ref{DFT}). The average and total variation of (\ref{maxdist}) read
\begin{eqnarray}
\label{maxdist2}
    \langle \Sigma \rangle' = \eta a' \tanh(a'/2),
    \\
\label{maxdist3}
    \Delta' = \eta \tanh(a'/2).
\end{eqnarray}
Using the average and total variation of the minimal distribution $p_a(\Sigma)$, we get $\langle \Sigma \rangle = a \tanh(a/2)$ and $\Delta^*=\tanh(a/2)$, which results in $\Delta^*=\tanh(g(\langle \Sigma \rangle)/2)=\langle \Sigma \rangle/a$.

Finally, define $k:=2\Delta^*/\epsilon>2$ and take $a':=ka$ and $\eta=h(a)/h(ka)$ in (\ref{maxdist}). Note that $0<\eta<1$, because $h(ka)>h(a)$, since $h'>0$ and $ka>a$. For this specific choice of $\eta$ and $a'$, we have $\langle \Sigma \rangle'= a \tanh(a/2)=\langle \Sigma \rangle$ from (\ref{maxdist2}), and from (\ref{maxdist3}) we obtain
\begin{eqnarray}
\label{maxdist4}
    \Delta' = \frac{\langle \Sigma \rangle'}{a'} = \frac{\langle \Sigma \rangle}{a'}=\frac{a\Delta^*}{a'}=\frac{\Delta^*}{k}= \frac{\epsilon}{2}< \epsilon.
\end{eqnarray}
Therefore $\Delta'<\epsilon$ and there is a distribution $p'(\Sigma)$ such that $P(\Sigma<0)+P_0/2\geq \frac{1}{2}(1-\epsilon)$, which contradicts inequality (\ref{ineq}) and proves that (\ref{upperbound}) is the best possible bound in this setup.

\begin{figure}[htp]
\includegraphics[width=3.3 in]{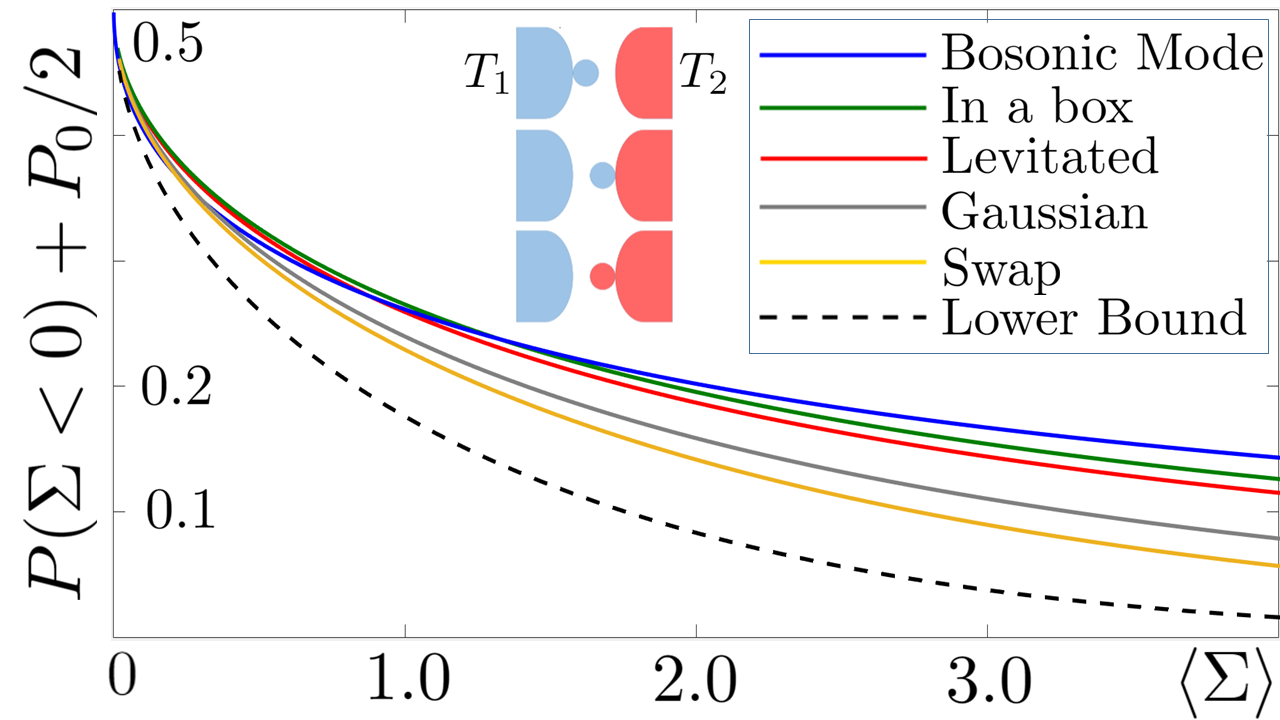}
\caption{(Color online) Apparent violation of the second law $P(\Sigma<0)+P_0/2$ as a function of the mean $\langle \Sigma \rangle$ for the nonequilibrium heat exchange problem mediated by different systems: a levitated nanoparticle (red), a particle in a box (green), a bosonic mode weakly coupled to the reservoirs (blue). For comparison, the apparent violation for a gaussian distribution (grey) and a qubit swap engine (yellow) are also depicted. All cases satisfy the DFT and present apparent violation above the lower bound (dashed), as expected. The classic examples (levitated, in a box and Gaussian) have $P_0=0$ for $\langle \Sigma \rangle >0$ and the quantum cases (bosonic mode and qubit swap engine) have some $P_0>0$. All examples (and the bound) collapse to $P(\Sigma<0)+P_0/2=P_0/2=1/2$ for $\langle \Sigma \rangle=0$ (equilibrium).}
\label{fig1}
\end{figure}

{\bf \emph{Application to a bosonic mode -}} We consider a free bosonic mode with Hamiltonian $H=\hbar\omega (a^\dagger a+1/2)$ weakly coupled to a thermal bath such that the density matrix satisfies a Lindblad's equation \cite{Santos2017a,Salazar2019,Denzler2019},
\begin{equation}
\label{Lind}
\partial_t \rho = \frac{-i}{\hbar}[H,\rho] + D_i(\rho),
\end{equation}
for the dissipator given by
\begin{equation}
D_i(\rho)=\gamma(\overline{n}_i+1)[a\rho a^\dagger - \frac{1}{2}\{a^\dagger a, \rho\}]+\gamma \overline{n}_i[a^\dagger \rho a -\frac{1}{2}\{a a^\dagger , \rho\}],
\end{equation}
where $\gamma$ is the dissipation constant and $\overline{n}_i=[\exp(\hbar \omega/k_BT_i)-1]^{-1}$ is the bosonic thermal occupation number and $\beta_i=1/(k_b T_i)$. At $t=0$, the system starts in thermal equilibrium (with temperature $T_1$) and it is placed in contact with the second reservoir (temperature $T_2$). The dynamics is then modeled by (\ref{Lind}) with $D_2$ as the dissipator for $t>0$. Let $\rho_t:=\Phi_t(\rho_0)$ define the dynamical map that solves (\ref{Lind}) with $D_2$. In a two point measurement scheme (TPM) at $t=0$ and $t>0$, it yields energy values $E_1$ and $E_2$, with energy variation $\Delta E = E_2-E_1$. 

Upon repeating the experiment several times, the energy variation distribution is given by $p(\Delta E=\hbar \omega m)=\sum_{n=0}^{\infty}\langle n+m | \Phi_t \big(|n\rangle \langle n |\big)|n+m\rangle p_n$, where $p_n={e^{-\beta_1 E_n}}/Z(\beta_1)$ and $Z(\beta_1)=\tr (e^{-\beta_1 H})$. It was showed that $p(\Delta E)$ has a closed form \cite{Salazar2019,Denzler2019}, so that the entropy production is given in terms of the energy variation \cite{Campisi2015,Timpanaro2019B,Sinitsyn2011} as $\Sigma = -(\beta_2-\beta_1)\Delta E$. In this case, the distribution $P(\Sigma)$ follows \cite{Salazar2019} directly:

\begin{equation}
\label{PsigmaHO}
    p(\Sigma)=\frac{1}{A(\alpha)} \exp(\frac{\Sigma}{2}-\alpha \frac{|\Sigma|}{2}),
\end{equation}
with support $s=\{\pm \Delta \beta \hbar \omega m\}=\{\pm\varepsilon m\}$, $m=0,1,2,..$, and constant $A=A(\alpha)$, for some constant $\alpha$. Note that (\ref{PsigmaHO}) satisfies (\ref{DFT}). Using the geometric series, the expression (\ref{lowerbound}) is computed exactly from (\ref{PsigmaHO}):
\begin{equation}
    \label{QHOviolation}
    P(\Sigma < 0 )+\frac{P_0}{2} = \frac{1}{2A(\alpha)}\frac{1+e^{-\gamma_{-}}}{1-e^{-\gamma_{-}}},
\end{equation}
where $\gamma_{\mp}=\varepsilon(\alpha\pm 1)/2$, with $\alpha$ and $\varepsilon$ defining $A$ and $\langle \Sigma \rangle$ uniquely from
\begin{eqnarray}
\label{QHOA}
A(\alpha)=\frac{1+e^{-\gamma_{+}}}{2(1-e^{-\gamma_{+}})}+\frac{1+e^{-\gamma_{-}}}{2(1-e^{-\gamma_{-}})},
\\
\label{QHOMean}
\langle \Sigma \rangle = \frac{\varepsilon}{A}
\big[\frac{e^{\gamma_{+}}}{(e^{\gamma_{+}}-1)^2}-\frac{e^{\gamma_{-}}}{(e^{\gamma_{-}}-1)^2}\big].
\end{eqnarray}

In order to test the bounds (\ref{lowerbound}) and (\ref{upperbound}), we select several $\alpha>0$ incrementally, with $\varepsilon=1$, then compute $A(\alpha)$ from (\ref{QHOA}) and use it to find and plot (\ref{QHOviolation}) vs. $\langle \Sigma \rangle$ from (\ref{QHOMean}) for each $\alpha$. The resulting curve is compared to the lower bound (\ref{lowerbound}) computed with the same $\langle \Sigma \rangle$, showed in Fig.1.

{\bf \emph{Application to a levitated nanoparticle -}} In the highly underdamped limit, the generalized Langevin equation represents the dynamics of a levitated nanoparticle \cite{Gieseler2012,Gieseler2018,Aspelmeyer2014,Salazar2019a} as well as a particle in a box \cite{Salazar2020,PhysRevLett.117.180603}. We start with the Langevin dynamics with a general single well potential $\mathcal{U}(x)=m k x^{2n}/2$. The particle's dynamics is given by
\begin{equation}
    \label{Langevin}
    \ddot{x} + \Gamma \dot{x} +\Omega_0^2 L \big(\frac{x}{L}\big)^{2n-1} = \frac{1}{m}F_{fluc}(t),
\end{equation}
for position $x(t)$, with Gaussian noise $\langle F_{fluc}(t)F_{fluc}(t')\rangle = 2m\Gamma T \delta(t-t')$, where $\Gamma$ is a friction coefficient, $m=1$ is the particle mass, $T$ is the reservoir temperature and $k=\Omega_0^2$ is a constant (not driven by a protocol). Defining the energy $E= p^2/2+\mathcal{U}(x)$ ($p=\dot{x}$), the following stochastic differential equation (SDE) was obtained for the total energy in the highly underdamped limit \cite{Gieseler2012,Salazar2019a}, $\Omega_0\gg\Gamma$:
\begin{equation}
\label{LangevinforE2}
dE=-\Gamma_n (E- \frac{f_n}{2}T)dt+\sqrt{2\Gamma_n TE}d\textrm{W}_t,
\end{equation}
with effective degrees of freedom $f_n=(n+1)/n$, friction coefficient $\Gamma_n=\Gamma [2n/(n+1)]$,  $d\textrm{W}_t$ is a Wiener increment. The particle is prepared with temperature $T_1$ and placed in contact with the reservoir $T_2$ for $t>0$, which is the same setup used with the bosonic mode. Again, one defines the entropy production as $\Sigma = -\Delta \beta \Delta E$, where $p(\Delta E)=\int P(E_1)\Pi_t(E_1\rightarrow E_2)\delta (\Delta E - (E_2-E_1))dE_1 dE_2$. The propagator $\Pi_t$ is known \cite{Gieseler2018,Salazar2016} for the SDE (\ref{LangevinforE2}), which yields the following distribution:
\begin{equation}
\label{nanopdf}
    p(\Sigma)=\frac{1}{B(\alpha)}
    \exp\big(\frac{\Sigma}{2}\big)|\Sigma|^{(f_n+1)/2}K_{(f_n+1)/2}(\alpha |\Sigma|),
\end{equation}
defined for the real line for constant $\alpha$ defined in terms of parameters ($T_1,T_2,\Gamma_n t$), $B(\alpha)$ is a normalization constant, and $K$ is the modified Bessel function of the second kind. In order to test the bounds (\ref{lowerbound}) and (\ref{upperbound}), we select some $\alpha>0$, then compute $P(\Sigma<0)$ and $\langle \Sigma \rangle$ numerically from (\ref{nanopdf}). In this continuous case, $P_0=0$. Finally, the resulting curve is compared with the lower bound (\ref{lowerbound}) computed with the same $\langle \Sigma \rangle$, showed in Fig.1 for the levitated case (harmonic, $n=1$) and particle in a box ($n\rightarrow \infty$).

{\bf \emph{Swap engine -}} As another example of entropy production pdf, take a pair of qubits with energy gaps $\epsilon_A$ and $\epsilon_B$. They are prepared in thermal equilibrium, $p(\pm)=\exp(\pm\beta \epsilon)/(\exp(-\beta\epsilon)+\exp(+\beta\epsilon))$, for $\beta\in\{\beta_1,\beta_2\}$ and $\epsilon\in\{\epsilon_A,\epsilon_B\}$, with reservoirs at temperature $T_1$ and $T_2$. A TPM is performed before and after a swap operation \cite{Campisi2015}, here defined as $|xy\rangle \rightarrow |yx\rangle$, for $x,y \in \{-,+\}$. The entropy production in the process is given \cite{Campisi2015,Timpanaro2019B} by $\Sigma = \beta_1 \Delta E_A + \beta_2 \Delta E_B$,
where $\Delta E_A = E_A^f-E_A^i$, $\Delta E_B=E_B^f-E_B^i$ are the variations of energy measurements before and after the swap. In this TPM, the three possible outcomes are $\Sigma \in s=\{0,\pm 2a\}$ for $2a=2(\beta_2\epsilon_B-\beta_1\epsilon_A)$. The distribution of $\Sigma$ is given by $p(\Sigma)=(1/Z_0)\exp(\Sigma/2)$, for $\Sigma \in s$, which satisfies the DFT (\ref{DFT}), and the apparent violation of the second law reads $P(\Sigma<0)+P_0/2=(\exp(-a)+1/2)/(1+\exp(a)+\exp(-a))$, where $\langle \Sigma \rangle=2a(\exp(a)-\exp(-a))/(\exp(a)+\exp(-a)+1)$ defines $a$ in terms of $\langle \Sigma \rangle$, also depicted in Fig.~1.

{\bf \emph{Gaussian case-}}
For completeness, we also compare the apparent violation of the second law with respect to a Gaussian distribution \cite{Pigolotti2017,Chun2019}.
\begin{equation}
\label{gaussian}
    p(\Sigma)=\frac{1}{2\sqrt{\pi \langle \Sigma \rangle}}\exp\Big(\frac{-(\Sigma-\langle \Sigma\rangle)^2}{4\langle \Sigma \rangle}\Big),
\end{equation}
In this continuous case, we have $P_0=0$ and $P(\Sigma < 0)$ is given in terms of the error function, also included in Fig.~1.

{\bf \emph{Discussion and Conclusions -}}
We have used the strong Detailed Fluctuation Theorem and a single constraint, $\langle \Sigma \rangle$, to find a lower bound for the apparent violation of the second law, $P(\Sigma < 0) + 1/2$, given by (\ref{introB}). The factor $P_0/2$ appears quantifying the apparent violation so it takes into account discrete domains of $\Sigma$. It is not the case for the minimal distribution and classical cases (as $P_0=0$), but it is relevant discrete situations ($P_0>0$, as showed in the quantum applications).

Intuitively, the bound is also given in terms of the minimal distribution, the same distribution that saturates the TUR (using the same constraint in $\langle \Sigma \rangle$).
In the applications, we showed the entropy production of different systems satisfying the lower bound as a function of $\langle \Sigma \rangle$. The behavior of the lower bound in Fig.~1 (dashed line) shows that it falls rapidly to zero for $\langle \Sigma \rangle\gg0$, indicating that the family of systems operating very far from equilibrium have enough room to prevent the second law to be violated. However, in situations close to equilibrium $\langle\Sigma\rangle \approx 0$, the lower bound imposes a high magnitude for the apparent violation of the second law, as observed in the systems considered here.

In situations where the DFT is valid, the area under the dashed curve in Fig.~1 is forbidden: meaning that one cannot devise a system that operates arbitrarily close to equilibrium ($\langle \Sigma \rangle\approx0$) without often ``violating'' the second law by at least the amount given by the lower bound.
The limitations of our result lie in the applicability of the strong DFT. For that reason, the results are relevant, for instance, in the context of exchange fluctuation theorems.
\bibliography{lib3}
\end{document}